\documentclass[amsmath,reprint]{revtex4-1} 

\usepackage{graphicx}  

\newcommand{\be}{\begin{equation}} 
\newcommand{\ee}{\end{equation}}
\newcommand{\bea}{\begin{eqnarray}} 
\newcommand{\eea}{\end{eqnarray}}

\begin{document}

\title{Phenomenological modeling of long range noncontact friction  in micro- and nanoresonators }

\author{Andr\'e Gusso} 
\email{andre.gusso@pq.cnpq.br}
\affiliation{Departamento de Ci\^encias Exatas-EEIMVR,
Universidade Federal Fluminense, Volta Redonda, 27255-125, RJ, Brazil.}

\date{\today}

\begin{abstract}
Motivated by the results of an experiment using atomic force microscopy performed by Gotsmann and Fuchs [Phys. Rev. Lett. {\bf 86}, 2597 (2001)], where a strong energy loss due to the tip-sample interaction was measured, we investigate the potential implications of this energy loss channel to the quality factor of suspended micro- and nanoresonators. Because the observed tip-sample dissipation  remains without a satisfactory theoretical explanation, two phenomenological models are proposed to generalize the experimental observations. A minimal phenomenological model simply extends for larger separations the range of validity of the power law found experimentally for the damping coefficient. A more elaborate phenomenological model assumes that the noncontact friction is a consequence of the Casimir force acting between the closely spaced surfaces. Both models provide quantitative results for the noncontact friction between any two objects which are then used to estimate the energy loss for suspended bar micro- and nanoresonators.  Its is concluded that the energy loss due  to the unknown mechanism has the potential to seriously restrict the quality factor of both micro- and nanoresonators. 
\end{abstract} 

\maketitle
\section{Introduction}

Noncontact friction, the energy loss due to the relative motion of bodies moving without contact,  has been receiving  great attention in the context of atomic force microscopy (AFM)\cite{Langewisch10}. It has been investigated  by its potential applications as a surface imaging technique, providing information about the surfaces not accessible by other techniques, as well as for the  understanding of the basic physical mechanisms underlying the dissipation process. Most of the recent theoretical and experimental efforts to understand this dissipation process were focused on the noncontact friction at very short tip-sample separations, typically below 1 nm. While still a matter of debate, in this short distance (large force) regime, the evidences suggest that the fundamental physical mechanism leading to energy dissipation  is force hysteresis\cite{Langewisch10}. Another important result regarding this subject was obtained by an earlier experiment which investigated the energy dissipation in AFM for comparatively larger tip-sample separations\cite{Gotsmann01}. More specifically, the dissipation between an aluminum coated silicon tip and a crystalline gold surface was measured and the results reported for tip-sample separation ranging from  contact up to 4 nm, for the case of a tip with a radius $R = 35 \pm 5$ nm, and 0.5 nm up to 7 nm for a tip with radius $R = 21 \pm 2$ nm.  Under the assumption of viscous damping, the noncontact friction for separations larger than 1.5 nm was characterized by  a distance-dependent damping coefficient which scaled with tip-sample separation $d$ as $\gamma(d) \propto d^{-3}$. In spite of the attempts to explain the observed dissipation, none of the known physical mechanisms leading to noncontact friction can explain the observed long range effects\cite{Gotsmann01,Volokitin07}. In fact,  the measured dissipation is usually orders of magnitude greater than any theoretical prediction \cite{Volokitin07}. The existence of such noncontact friction for the aluminum-gold system rises questions about its potential implications for a variety of micro- and nanosystems or devices. The same large dissipation could exist for other material combinations, and its effects could extend over larger separations between material bodies.

In the present work the dissipation due to an energy loss mechanism (ELM) described by two phenomenological models intended to generalize the results presented by Gotsmann and Fuchs\cite{Gotsmann01} (GF) is investigated for rectangular cross section suspended beam micro- and nanoresonators. The loss of vibratory mechanical energy is one of the most important problems to be addressed into the current development of micro- and nanoresonators, specially for microelectromechanical (MEM) and nanoelectromechanical (NEM) resonators. Several anelastic processes and clamping, for example, are known sources of energy dissipation\cite{LossMechanism,Cleland} and have been investigated both theoretically and experimentally. However, it was only recently that noncontact friction was considered as a source of energy dissipation in such devices\cite{Gusso10a,Gusso10b}. In general, the noncontact friction for suspended micro- and nanoresonators arises from the interaction of the resonator with surrounding structures over vacuum or air gaps. Its  relevance is evidenced, for instance, by the results presented in Ref. \onlinecite{Gusso10b} where, as a consequence of acoustic electromechanical energy loss,  low quality factors are predicted for realistic microresonators. 

Since no fundamental physical mechanism is known to explain the results of GF, the proposed ELM for the resonators investigated in this work is motivated by the analogy between the experimental setup of GF and a suspended resonator separated from nearby structures through nanogaps. In both cases we have surfaces in close proximity and in a periodic relative motion. Therefore, dissipation of the resonator mechanical energy can be expected to be induced by the same mechanism present at the tip-sample system in the typical  configuration considered here, where the resonator oscillates perpendicular to an underlaying plane surface (for example, the substrate or large electrodes). In principle, this analogy is strictly valid when the resonator and the nearby plane surface are made from the same materials used in  the GF experimental tip-sample system. However, because the number o possible physical phenomena in any simple system is limited, we can expect to observe the same physical mechanism inducing  noncontact friction in analogous systems involving other materials. In fact, at least for small tip-sample separations, energy dissipation of the same order of magnitude has been observed for other tip-sample materials combination involving metals, semiconductors and insulators\cite{Langewisch10,Roll08,Gotsmann99}. It is this possibility that makes the present investigation more relevant, since the results and conclusions can be valid for other systems of practical interest. 

The article is organized as follows. In section \ref{Qualityfactor} we derive a suitable expression for the quality factor $Q$ in terms of an areal damping coefficient. In section \ref{models} two phenomenological models intended to explain and/or extend the range of applicability of experimental results and their predictions for $Q$ are presented. Further discussions and our conclusions are presented in section \ref{conclusion}.

\section{Quality Factor}
\label{Qualityfactor}

In this section we derive the expression for the quality factor of a rectangular cross section suspended resonator which loses energy due to a viscous damping force $F^{viscous} = \gamma(d) v$, where $\gamma(d)$ is the damping coefficient.  The resonator is considered to be homogeneous, having length $l$, thickness $h$, and width $w$, each dimension being parallel to the $x, y$ and $z$ axis, respectively. It can be clamped or free at each end, resulting in a cantilever, bridge or free resonator that vibrates transversally in the $z$ direction. It vibrates close to a plane surface contained in the $xy$ plane and situated at the average distance $d_0$. The mode shapes are those given by the Euler-Bernoulli beam theory, namely,    $u_n(x) = u_0 \{ \cosh(\kappa_n x/l) -\cos(\kappa_n x/l) + \chi_n [\sinh(\kappa_n x/l) -$ $\sin(\kappa_n x/l) ] \}$, where $n$ is the mode index ($n=1,2,3...$), and $\kappa_n$ and $\chi_n$ are determined accordingly for each boundary condition\cite{Cleland}. 

The quality factor is defined as $Q= 2 \pi U_n/\Delta U_n = \omega_n U_n/\Pi$, where $U_n$ denotes the total mechanical energy stored into the resonator, and $\Delta U_n$ corresponds to the energy lost per cycle, which can be calculated from the time averaged energy loss $\Pi$. The vibrational energy  for  cantilevers, bridges and free resonators can be very well approximated by $U_n = h w l \rho \omega_n^2 u_0^2/2$, where $\rho$ denotes the density and $\omega_n$ the mode frequency given by $\omega_n^2 = \kappa_n^4 E h^2/(12 \rho l^4)$, with $E$ the Young modulus. The only missing ingredient for the calculation of $Q$ is the average energy loss. As we discuss further in the next section, we are going to assume that the noncontact friction is characterized by a distance dependent areal damping coefficient $\Gamma(d)$. In such a case, a resonator vibrating with a displacement $u_n(x,t) = u_n(x) \sin(\omega_n t)$ dissipates the power,
\bea
P(t) &=& \int_0^l \Gamma[d(x)]\, \dot{u}_n(x,t)^2 \,w\, dx \nonumber \\ 
&\approx& \Gamma(d_0)\, w \,\omega_n^2 \int_0^l u_n(x)^2 dx \; \sin^2(\omega_n t) , 
\eea
where the approximate expression is obtained by taking $d(x)=d_0+u_n(x) \approx d_0$, because of the small amplitude of oscillation required for operation in the linear regime. 

Taking into account the orthonormality of the mode shapes $u_n(x)$ the average dissipated power results to be
\be 
\Pi = \frac{1}{2} \Gamma(d_0) \, w \, l \, \omega_n^2 u_0^2 ,
\ee
leading to the quality factor
\be
Q = \frac{\kappa_n^2}{\sqrt{12}} \frac{\sqrt{\rho E}}{\Gamma(d_0)}\left(\frac{h}{l} \right)^2 .
\label{Q}
\ee
The resulting quality factor depends on the geometry of the resonator only through the ratio $h/l$. It does not depend on the actual dimensions of the resonator, being the same for both micro- and nanoresonators.

\section{Phenomenological models and results}
\label{models}

\subsection{Minimal phenomenological model}

\subsubsection{Model}

In order to be able to use the experimental results for dissipation measured for a sphere-plane geometry (used to model the tip-sample system), for the approximately plane-plane geometry of the vibrating resonator and the nearby substrate, a fundamental assumption is that a physically meaningful areal damping coefficient $\Gamma(d)$ can be defined. By physically meaningful it is meant that $\Gamma(d)$ is the fundamental physical quantity determining the overall damping coefficient. $\gamma(d)$ is then obtained by performing a suitable integration limited to the overlapping area of the two surfaces. We  note that for several known  noncontact friction mechanisms a physically meaningful $\Gamma(d)$ can, in fact, be defined. That is the case for the van der Waals (vacuum) friction\cite{Volokitin07} or stochastic friction\cite{Gauthier99}. However, for the electrostatic friction only  $\gamma(d)$ can be defined due to the electric field dependence on the overall surface geometry, and similarly for the phonon emission mechanism, where the surface stress is, in general, a complex function of the applied forces\cite{Volokitin07,Gusso10a,Gusso10b}. 

From the experimental data obtained by GF for the tip-sample damping coefficient $\gamma(d)$ it is straightforward to determine $\Gamma(d)$.  First, we note that for any two interacting surfaces the general relation between  $\gamma$ and $\Gamma$ is 
\be 
\gamma(d) = \int \Gamma \, dA ,
\label{gammaint}
\ee
where the integral is performed over the overlapping area, and $d$ is a conveniently defined distance between the surfaces. In general, it is not straightforward to determine $\Gamma(d)$  from this relation, however, in the case of the tip-sample system we can resort to the Deryaguin approximation\cite{Hunter01}. We note that all the mathematical requirements for the use of  the Deryaguin approximation in the evaluation of Eq. (\ref{gammaint}) are satisfied, namely,  $\Gamma(d)$   is a fast decreasing function of surface separation (based on fact that $\gamma(d) \propto d^{-3}$), and  the tip has a radius of curvature $R$ satisfying $R \gg d$, where $d$ denotes the distance of closest sphere-plane separation. Using the Deryaguin approximation the approximate relation
\be
\gamma(d) = 2 \pi R \int_d^\infty \Gamma(z) dz .
\label{gG}
\ee
can be obtained, where $z$ is the vertical coordinate perpendicular to the plane surface. Deriving Eq. (\ref{gG}) with respect to $d$  allow us to get,
\be
\Gamma(d) = -\frac{1}{2 \pi R} \frac{\partial \gamma(d)}{\partial d}.
\label{gammad}
\ee
It has to be noted that it is due to the particular sphere-plane geometry in the experiment of GF that we can easily obtain the more fundamental physical quantity $\Gamma(d)$ from the experimental results on  $\gamma(d)$. 

From the fitting of the experimental data on the average dissipated energy (Figure 3 of Ref. \onlinecite{Gotsmann01}) as a function of the distance of closest tip-sample separation GF provided an expression for $\gamma(d)$ valid in the range $1.5 \lesssim d \lesssim 7.0$ nm
\be
\gamma(d) = \frac{\gamma^0}{d^3} = \frac{(8.0^{+5.5}_{-4.5})\times 10^{-35}}{d^3} ,
\label{fitGF}
\ee
where the approximate range of validity results from the uncertainties on the definition of the tip-sample distance inherent to the AFM technique. We can use this result and Eq. (\ref{gammad}) to calculate $\Gamma(d)$ in the range where the fit applies. However, the currently developed micro- and nanoresonators are made with gaps larger than approximately  20 nm (see Ref. \onlinecite{Agache05} for the smallest gap produced so far), and a model to extend the results of GF to larger distances becomes necessary in order to estimate the resulting quality factor for $d \gtrsim 7.0$ nm. 

Based on the fact that $\Gamma(d)$ or $\gamma(d)$ predicted by some of the known noncontact friction mechanisms also obey simple power laws on the variable $d$, it is reasonable to assume that the power law  for $\gamma(d)$ evidenced by the data fitting reflects the existence of an underlying physical mechanism that would be operating over larger distances. For this reason, we consider a minimal phenomenological model (MPM) which predicts $\Gamma(d)$ by the substitution of $\gamma(d)$ into Eq. (\ref{gammad}). The resulting damping coefficient is
\be
\Gamma(d) = \frac{3 \gamma^0}{2 \pi R}\frac{1}{d^4} = \frac{\Gamma^0}{d^4},
\ee
and is assumed to be valid for $d \gtrsim 1.5$ nm. $\Gamma^0$ can be calculated taking into account the errors on the determination of $\gamma^0 = (8.0^{+5.5}_{-4.5})\times 10^{-35}$ Nsm$^2$ and $R = 21 \pm 2$ nm. For an estimate of $\Gamma^0$ we assume symmetrical errors for $\gamma^0$ equal to $\pm 5 \times 10^{-35}$ Nsm$^2$ which results into $\Gamma^0 = (1.8 \pm 1.2) \times 10^{-27}$  Nsm. 

We have just determined $\Gamma(d)$ from measurements performed for a system comprised of two interacting surfaces, one being crystalline gold and the other a not fully characterized aluminum surface\cite{Gotsmann01}. While the results for $\Gamma(d)$ can be considered as strictly valid for aluminum-gold systems, the analysis of the experimental data for a silicon-mica system performed in section \ref{conclusion} indicates that the same ELM can exist for other material combinations. Assuming this point of view, we consider that $\Gamma(d)$ obtained for the aluminum-gold system is only a reference value. In spite of the fact that no systematic experimental results obtained so far can exclude the role of the ELM described by the MPM on any other two materials combination, we can still be conservative and consider that this ELM is restricted to metallic systems. In this case, our results for $Q$ would only be valid for whole metallic resonators\cite{Li07} or those, for instance, having metalized electrodes\cite{Wang05}. However, in what follows, no specific assumptions are made with regard to the exact structure of the resonator and the underlying substrate, and for the purpose of the mechanical modeling the resonator is assumed to have a homogeneous and purely elastic structure. 
    
\subsubsection{Results} 
\label{ResultMPM}   
    
\begin{figure}
\includegraphics{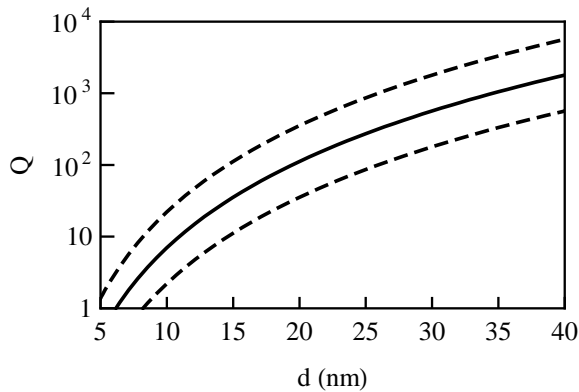}
\caption{\label{Qsimplefig} Log-linear plot of the quality factor $Q$ predicted by the MPM as a function of the gap $d$ for a doubly clamped polysilicon resonator with aspect ratio $h/l=0.1$. The continuous line corresponds to $\Gamma^0 = 1.8 \times 10^{-27}$ Nsm. The upper(lower) dashed lines are obtained by dividing(multiplying) $\Gamma^0$ by $\sqrt{10}$.} 
\end{figure}

In Fig. \ref{Qsimplefig} we present a range of $Q$ values as a function of the gap $d$, for a doubly clamped polysilicon ($E = 170$ GPa and $\rho = 2.3 \times 10^3$ kg)  resonator having a typical aspect ratio $h/l = 0.1$. We let $\Gamma^0$, and consequently $Q$, vary over one order of magnitude around the reference value $\Gamma^0 = 1.8 \times 10^{-27}$ Nsm. We have chosen polysilicon as a reference material because it is the most frequently used structural material in micro- and nanodevices.   Values of $Q$ in the same range are predicted for other structural materials  because of the relatively small range of variation (within a factor of two) of the factor $\sqrt{\rho E}$ appearing in the Eq. (\ref{Q}) for materials of relevance in micro- and nanofabrication including metals, semiconductors, and insulators.

The predicted $Q$ is rather small, specially below 20 nm, even for the smallest value assumed for $\Gamma^0$. Such quality factors, well below $10^3$, are smaller than some of the lower values obtained experimentally for suspended bar nanoresonators not under the influence of nearby surfaces\cite{Li07, Huang05}. Therefore, this ELM has the potential to significantly degrade $Q$ if the currently fabricated nanoresonators are, for instance, set to operate as electromechanical devices actuated electrostatically through nanometric gaps. It has also to be noted that, while the aspect ratio $h/l=0.1$ is typical, much smaller aspect ratios can be easily found for practical devices, implying that the impact of this ELM on the degradation of $Q$ can be even more significant. This ELM can also be relevant for MEM resonators, currently fabricated with gaps as small as about 30 nm\cite{Wang00}, because the predicted $Q$ is smaller than the normally found intrinsic quality factor  of the order of $10^4$. Another important result, which does not depend on the validity of the proposed MPM but, instead, depends solely on the  validity of the assumption that there is a $\Gamma(d)$ given by Eq. (\ref{gammad}), is the prediction of extremely small $Q$ in the range supported by the experimental data $d \lesssim 7$ nm. This result indicates that noncontact friction, as measured by GF, can pose stringent limits on the quality factor of small gap nanoresonators or, equivalently, may set limits to the smallest possible practical gaps.

\subsection{Extended phenomenological model}
\label{EPM}

\subsubsection{Model}
\label{EPMmodel}

The results obtained based on the MPM demonstrate the potential relevance of the noncontact friction mechanism unveiled by GF to both suspended micro- and nanoresonators. Therefore, a better understanding of this noncontact friction mechanism is desirable. In the search for a phenomenological model that could account for the experimental data presented by GF in a simple an coherent form we first note that the tip-sample force $F^{ts}$ is essential in determining the friction resulting from different mechanisms such as phonon emission\cite{Volokitin07}, stochastic friction\cite{Gauthier99}, viscoelastic dissipation\cite{Boisgard02}, and the simple model based on two-level systems of Ref. \onlinecite{Hoffmann01}. In spite of the fact that none of those mechanism can explain the dissipation observed by GF they evidence the possible relation between $F^{ts}$ and noncontact friction.  The functional dependence of dissipation on the force can be quite involved as for the viscoelastic dissipation\cite{Boisgard02}. It can also be straightforward as is the case for stochastic friction and the two-level systems mechanism proposed in Ref. \onlinecite{Hoffmann01}. In the first case it is predicted that $\gamma(d) \propto (\nabla F^{ts})^2$, while the last mechanism  predicts $\gamma(d) \propto F^{ts}$.  

Considering that in the experiment of GF the role of electrostatic force was minimized living  the Casimir force\cite{Bordag01} as the only relevant force at large tip-sample separations,  we further analyzed  the data presented by GF and established a possible  relation between the tip-sample force and the damping coefficient. We note that $\gamma(d)$ given in Eq. (\ref{fitGF}) depends on $d$ as $d^{-3}$ in a distance range where the Casimir force reduces to the van der Waals force of the form  $F^{vdW} = -H R/(6 d^2)$, where $H$ denotes the Hamaker constant\cite{Hunter01}, and whose gradient $\nabla F^{ts} = \partial F^{ts}/\partial d = H R/ (3 d^3)$ also goes as $d^{-3}$. This fact suggests the possibility  that  $\gamma(d)$ be proportional to  the force gradient. Based on this possibility we formulate the following hypothesis for the relation between the tip-sample force and the damping coefficient
\be
\gamma(d) =  C \frac{\partial F^{ts}}{\partial d},
\label{hyp}
\ee
where $C$ is a phenomenological constant.

\begin{figure}
\includegraphics{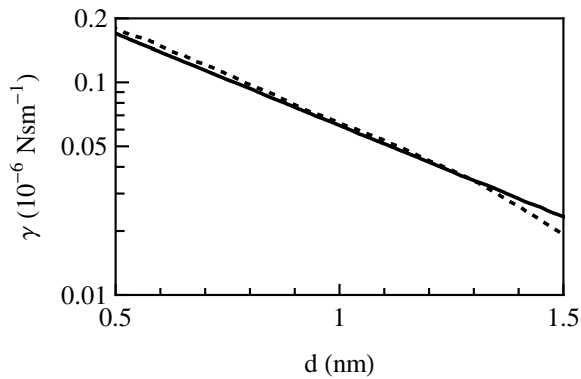}
\caption{\label{gammafig} Log-linear plot of the experimental $\gamma(d)$ as extracted from Fig. 1 of GF (continuous line), and the theoretical $\gamma(d)$ calculated from the experimental data for the tip-sample force and Eq. (\ref{hyp}) (dashed line).}
\end{figure}

Unfortunately, the data provided by GF are presented for experiments involving AFM tips with different radius and do not form a coherent data set that allow us to test our hypothesis over a broader range of probed tip-sample distances. The data for the energy dissipation presented in Fig. 3 of Ref. \onlinecite{Gotsmann01} in the range 0.5-7.0 nm , and fitted in the range 1.5-7.0 nm leading to Eq. (\ref{fitGF}), is not accompanied by the corresponding data on the force that could be used to calculate the energy dissipation predicted based on the hypothesis Eq. (\ref{hyp}).  However,  another data set for  the measured tip-sample force and the corresponding $\gamma(d)^{exp}$ (obtained from a special numerical fitting of the experimental data on the dissipated power\cite{Gotsmann01}), referring to measurements performed using a tip with $R = 35 \pm 5$ nm, was presented by GF in Fig. 1. This figure provides the curves that allow for a test of the hypothesis we formulated in the range of short distances below 1.5 nm where the force increases exponentially. The data obtained from the curves using a high resolution automated digitizer is not sufficiently precise for an analysis for $d \gtrsim 1.5$ nm, where the force starts its transition to pure van der Waals force.  From the extracted data points for the experimental force we  numerically calculate the force gradient, the data being smoothed by the method of moving averages. By comparison with the curve for $\gamma(d)^{exp}$ we determined the value of the phenomenological constant as $C^{Al-Au} = (1.1 \pm 0.7) \times 10^{-8}$ s, where the error was estimated from the combined error in the calculation of $C = \gamma(d)^{exp}/\nabla F^{ts}$, with the uncertainties on $\gamma(d)^{exp}$ and the force being taken as 40$\%$\cite{Gotsmann01}. The experimental and theoretical curves can be seen in Fig. \ref{gammafig}, where  the results are presented in the range 0.5-1.5 nm. For $d < 0.5$ nm $\gamma(d)^{exp}$ keeps growing exponentially while the theoretical $\gamma(d)$ tends to decrease.  The good superposition of the experimental and theoretical curves serves to corroborate the hypothesis represented by Eq. (\ref{hyp}), which seems to be valid for $d \gtrsim 0.5$ nm. Below 0.5 nm some other noncontact friction mechanism that does not complies  with our hypothesis can dominate the dissipation.

Motivated by the seeming validity of the hypothesis represented by Eq. (\ref{hyp}), we propose a second phenomenological model that not only explains the measured dissipation in the range $0.5 < d < 7.0$ nm but is also assumed to be valid for larger distances. The extended phenomenological model (EPM) is then based  upon the previous assumption that $\gamma(d)$ derives from  the more fundamental physical quantity $\Gamma(d)$ with the addition of the new hypothesis. Now, taking into account that Eq. (\ref{hyp}), so far, is considered strictly valid for a tip-sample system, we can apply the Deryaguin approximation to rewrite both $\gamma(d)$ and $F^{ts}$ in such a way that
\bea
\gamma(d) &=&  C \frac{\partial F^{ts}}{\partial d} \nonumber \\ 
 \Rightarrow 2 \pi R \int_d^\infty \Gamma(z) dz &=&
2 \pi R C \frac{\partial}{\partial d} \int_d^\infty F^{ss}(z) dz ,
\eea
where $F^{ss}(z)$ denotes the  pressure between two flat surfaces separated by a distance $z$. Consequently, $\Gamma(d)$ and $F^{ss}(d)$ must be related by
\be
\Gamma(d) =  C \frac{\partial F^{ss}(d)}{\partial d}.
\label{hyp2}
\ee
Before we proceed, we note that this fundamental relation, obtained by performing an approximate calculation to the particular tip-sample system, could be derived more rigorously if the hypothesis expressed in Eq. (\ref{hyp}) was generalized by stating that the damping coefficient is directly proportional to the force gradient for any two interacting surfaces. This is a reasonable but not necessary generalization, and we do not discuss it further.

\subsubsection{Results}

To calculate $Q$ predicted by the EPM we need to know $C$, and determine $F^{ss}(d)$ for our specific system. Because we are interested in systems with gaps larger than a few nanometers, $F^{ss}(d)$ is dominated by the Casimir force as predicted by the Lifshitz theory\cite{Bordag01,Gusso08}. This force results from the quantum fluctuations of the vacuum electromagnetic field, and is a function of the optical properties of the surfaces. It can be conveniently expressed as
\be
F^{ss}(d) = \eta(d) \frac{\hbar c \pi^2}{240 d^4}= \eta(d) F^C(d) ,
\ee
where $\eta(d)$ corresponds to the correction factor to the Casimir force between two perfectly conducting surfaces $F^C(d)$. In the present analysis, this correction factor takes into account only the effects of the finite conductivity of the Au and Al surfaces, and was calculated using the Lifshitz theory as described, for instance, in Ref. \onlinecite{Gusso08}.  The optical data for both metals is that provided by Palik\cite{Palik85}. At small separations of a few nanometers  $\eta(d) \propto d$, and the Casimir force reduces to the van der Waals force. Having obtained  Hamaker constant, $H^{Al-Au} = 3.8 \times 10^{-19}$ J, we can calculate the force between two plane surfaces and determine $C$ in an alternative manner from the same data used to evaluate $Q$ predicted by the MPM. Substituting  $F^{ts}(d) = F^{vdW}(d)$ in Eq. (\ref{hyp}), and taking the experimental result for $\gamma(d)$ from Eq. (\ref{fitGF}), valid for the short distance limit,  we obtain $C^{Al-Au} = (3.0^{+2.6}_{-1.8})\times 10^{-8}$ s, where the errors derive from the uncertainties on $\gamma^0$ and $R$. While the central value differs significantly from the previously obtained  value of $C^{Al-Au} = (1.1 \pm 0.7) \times 10^{-8}$ s, both results can be reconciled  if we take into account the estimates for the errors. 

\begin{figure}
\includegraphics{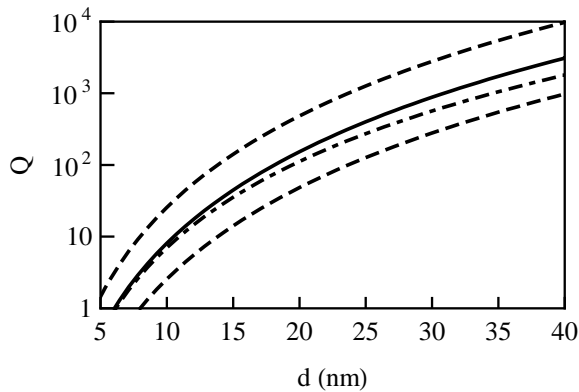}
\caption{\label{Qextendedfig} Log-linear plot of the quality factor $Q$ predicted by the EPM as a function of the gap $d$ for a doubly clamped polysilicon resonator with aspect ratio $h/l=0.1$. The continuous line corresponds to $C^{Al-Au} = 3.0 \times 10^{-8}$ s. The upper(lower) dashed lines are obtained by dividing(multiplying) $C^{Al-Au}$ by $\sqrt{10}$. The dot-dashed line presents $Q$ predicted by the MPM and corresponds to the continuous line in Fig. \ref{Qsimplefig}.} 
\end{figure}

Because the value $C^{Al-Au} = (3.0^{+2.6}_{-1.8})\times 10^{-8}$ s was determined from the same $\gamma(d)$ used to calculate  $\Gamma(d)$  predicted by the MPM, an adequate comparison between the predictions of both MPM and EPM requires the adoption of this value for $C$ and not the value obtained from other data set. For the sake of comparison  $Q$ predicted by the EPM was calculated for the same resonator considered for the MPM using $\Gamma(d)$ given in Eq. (\ref{hyp2}) with $F^{ss}(d)$ calculated numerically from Lifshitz theory, and using as a reference value $C^{Al-Au} = 3.0 \times 10^{-8}$ s. The results gained from the EPM are compared with those predicted by the MPM in Fig. \ref{Qextendedfig}. As expected from the faster decrease of the force predicted by the Lifshitz theory when compared to the van der Waals force, the dissipation predicted by the EPM becomes smaller than that for the MPM for larger distances resulting into higher quality factors. It can be seen that because the Casimir force reduces to the van der Waals force varying as $d^{-3}$ at short separations the predictions for $Q$  coincide for small $d$.

At this point, it is worth to compare our results with those obtained from another phenomenological model of noncontact dissipation proposed earlier by Dedkov\cite{Dedkov05}. The model is proposed in an attempt to get an unified explanation for the experimental results for both parallel and transversal relative motion between tip and sample. This model assumes that the damping coefficient is directly proportional to the tip-sample force instead of the force gradient, that means $\gamma(d) = V^{-1} F^{ts}(d)$, where $V$ is a phenomenological velocity to be determined from the experimental data.  The model was tested using the data on tip-sample force and damping coefficient for the Al-Au system investigated by GF and for the silicon-mica system investigated in Ref. \onlinecite{Gotsmann99}, in the case of relative transversal motion.  The analysis was restricted to short tip-sample separations, smaller than 2 nm, and $\gamma(d)$ predicted based on the experimental data on the tip-sample force agreed very well with the experimental data on $\gamma(d)$ with an adequate choice of the phenomenological velocity. As we discuss further in the next section, the predictions of the EPM for $\gamma(d)$ using the experimental data on the force for the silicon-mica system are also in good agreement with the measured $\gamma(d)$. The reason both phenomenological models can account for the experimental data can be attributed to the fact that the measured $F^{ts}(d)$ and $\gamma(d)$ are exponentially varying  functions of $d$ in the range $0.5 \lesssim d \lesssim 2.0$ nm. The EPM nevertheless can also account for the experimental results for $d > 2$ nm, where the phenomenological model proposed by Dedkov predicts incorrectly that $\gamma(d) \propto d^{-2}$ for the Al-Au system.  

\section{Discussion and Conclusions}
\label{conclusion}

\begin{figure}
\includegraphics{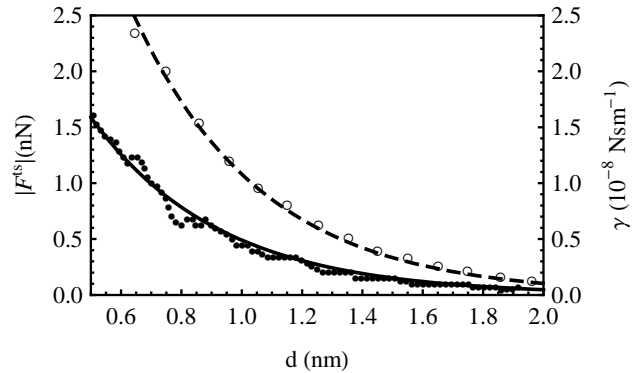}
\caption{\label{Simicafig} Data points and best fit curves ($\beta = 2.35$ nm$^{-1}$) for $|F^{ts}|$ ($\bullet$/continuous line) and $\gamma$ ($\circ$/dashed line) for the silicon-mica system. } 
\end{figure}

The results presented so far for $Q$ based on the minimal and extended phenomenological models highlight the potential relevance of an yet unknown energy loss mechanism on the performance of micro and nanoresonators. However, the determination of the phenomenological constant and the corroboration of the EPM relied on experimental results for an Al-Au system, therefore restricting the generalization for other systems of practical interest, as mentioned at the Introduction. Fortunately, the experimental results on force and dissipation for a semiconductor-insulator system is also available that further corroborate the EPM. In Ref. \onlinecite{Gotsmann99} the tip-sample force and dissipation were measured for a silicon tip ($R \approx 10$ nm) vibrating close to a mica surface,  the data presented in the Fig. 7 of Ref. \onlinecite{Gotsmann99} for $F^{ts}(d)$ and $\gamma(d)$ being accurate for a comparison between theory and experiment for tip-sample separations limited to less than approximately 2 nm. The data points for the force were extracted using an automated digitizer and are presented in Fig. \ref{Simicafig}. The experimental data for the damping coefficient was conveniently extracted from the work of Dedkov\cite{Dedkov05}, where it is reproduced, and is also presented in Fig. \ref{Simicafig}. As for the Al-Au system in the short range limit, the best fit for both force and damping coefficient for $d$ in the range 0.5-2.0 nm were obtained with exponentially varying functions of the form $\alpha \exp(-\beta d)$, where $\alpha$ and $\beta$ are the fitting parameters. While the best fit value of $\beta$ for $F^{ts}(d)$ and $\gamma(d)$ differ by about $10\%$, adequate goodness of fit and predictions differing from the best fit values by less than the estimated experimental errors are obtained by the assumption, consistent with the EPM, of an intermediate value of $\beta = 2.35$ nm$^{-1}$ for both $F^{ts}(d)$ and $\gamma(d)$. The resulting best fit curves are plotted in Fig. \ref{Simicafig}.

The phenomenological constant obtained from the substitution of the best fit functions in Eq. (\ref{hyp}) was $C^{Si-mica} = 0.94 \times 10^{-8}$ s, a result comparable to that for the Al-Au system for the short-range data. More precisely, performing the same analysis performed above for the silicon-mica system the best fit to the short-range data for the Al-Au system results in an inverse decay length $\beta = 2.2$ nm$^{-1}$, similar to that for the silicon-mica system. Furthermore, the phenomenological constant obtained by this procedure results to be $C^{Al-Au} = 1.3 \times 10^{-8}$ s, a result compatible with that derived directly from the experimental data by the curve matching in section \ref{EPMmodel}. 

From this discussion we can conclude that the EPM is corroborated by experiments involving very distinct materials combination. Additionally, the phenomenological constant that characterizes the yet unknown underlying physical mechanism has similar values for both materials combination, indicating that the fundamental process leading to energy dissipation could be the same irrespective to the materials involved. However, further experimental investigations of noncontact friction are required, specially for larger separations, because several questions are still open regarding, for instance, the role of  specific physical conditions, chemical composition, roughness, temperature and geometry of the surfaces. 

The experimental results should provide a guidance for more fruitful theoretical investigations that goes beyond the phenomenological models.  In fact, we note that the results from a set of five experiments already point toward a fundamental role of surface imperfections\cite{Langewisch10,Gotsmann01,Gotsmann99,Stipe01,Kisiel11} on energy dissipation. All five experiments involved the interaction of an imperfect tip surface with, nominally perfect, atomically flat surfaces. In the three experiments where the tip vibrates perpendicular to the sample surface\cite{Langewisch10,Gotsmann01,Gotsmann99} and, consequently, is subject to a time varying force  damping  is several orders of magnitude larger than that for the two experiments\cite{Stipe01,Kisiel11} where the tip vibrates parallel to the sample and is subject to a constant force. Because in both configurations the sample surface is always subject to a time varying force it is reasonable to conclude that the main contribution to the very distinct damping levels  comes from energy dissipation due to the time varying force acting at the tip surface. In fact, the experiment by Kisiel {\it et al.}\cite{Kisiel11}, has already elucidated that in the case of parallel oscillation the main contribution to noncontact friction comes from electronic friction, a mechanism that can not explain the dissipation for vertical oscillations.  The role of the tip surface  is further corroborated by the fact that dissipation has shown no dependence on  sample temperature in the experiment reported in Ref. \onlinecite{Langewisch10}, a result that alone has led the authors of this reference to consider that  the major contribution (to dissipation) stems from mechanisms within the tip\cite{Langewisch10}.  It has to be noted that in this last work experimental results for the tip-sample force and dissipation are also presented for  silicon tips on crystalline NaCl(001) surface. However, the range of useful data for analysis ranges from 0.5 nm up to only approximatelly 1.2 nm and leads to results that are inconclusive with regard to the relation between $F^{ts}(d)$ and $\gamma(d)$. Such a result is expected if the tip radius becomes too small compared to $R \approx 20-35$ nm in the experiment by GF, because for small tip radius the role of the atoms at the tip apex can dominate the overall tip-sample force and dissipation\cite{Ghasemi08} at the very small tip-sample separations where they are experimentally accessible. In this case the prevailing mechanism of energy dissipation  seems to be force hysteresis as evidenced by atomistic simulations\cite{Ghasemi08} and experimental results\cite{Schirmeisen05,Oyabu06}.        

Finally, from the results for the quality factors predicted based on both the MPM and the EPM, we conclude that the yet unknown ELM  first observed by GF can pose stringent limits on two of the most relevant operational parameters of both micro- and nanoresonators, the quality factor and the minimum gap.   For instance, because low mechanical impedances and high $Q$ are usually sought in MEM and NEM resonators for RF applications, a trade off between these two operational parameters may be a consequence of this ELM, for the impedance  and $Q$ increase rapidly with the  gap. Further experimental investigations aimed at measuring dissipation at larger separations  up to tens of nanometers should provide the pieces of information required  to test the two phenomenological models, and for gaining a better understanding of the long range noncontact friction  and its consequences for micro- and nanodevices.

\acknowledgements

The author wish to thank D. C. Lob\~{a}o and R. M. de Almeida for the useful conversations on data fitting, and the authors of Ref. \onlinecite{Langewisch10} for kindly providing the experimental data reported in their work.
This work was supported  by the  Conselho Nacional de Desenvolvimento Cient\'ifico e Tecnol\'ogico, CNPq-Brazil.

\end{document}